\documentclass{article}
\usepackage{eurosym}
\usepackage{amssymb}
\usepackage{amsmath}
\usepackage{cite}

\newtheorem{theorem}{Theorem}

\newtheorem{proposition}[theorem]{Proposition}

\begin{document}

\title{Symmetry analysis for the $2+1$ generalized quantum
Zakharov-Kuznetsov equation}
\author{Andronikos Paliathanasis\thanks{%
Email: anpaliat@phys.uoa.gr} \\
{\ \textit{Institute of Systems Science, Durban University of Technology }}\\
{\ \textit{PO Box 1334, Durban 4000, Republic of South Africa}} \\
{\textit{Instituto de Ciencias F\'{\i}sicas y Matem\'{a}ticas,}}\\
{\ \textit{Universidad Austral de Chile, Valdivia, Chile}}\\
\and P.G.L. Leach \\
{\ \textit{Institute of Systems Science, Durban University of Technology }}\\
{\ \textit{PO Box 1334, Durban 4000, Republic of South Africa}} \\
{\ \textit{School of Mathematical Sciences, University of KwaZulu-Natal }}\\
{\ \textit{Durban, Republic of South Africa}}}
\maketitle

\begin{abstract}
We solve the group classification problem for the $2+1$ generalized quantum
Zakharov-Kuznetsov equation. Particularly we consider the generalized
equation $u_{t}+f\left( u\right) u_{z}+u_{zzz}+u_{xxz}=0$, and the
time-dependent Zakharov-Kuznetsov equation $u_{t}+\delta \left( t\right)
uu_{z}+\lambda \left( t\right) u_{zzz}+\varepsilon \left( t\right) u_{xxz}=0$%
. Function $f\left( u\right) $ and $\delta \left( t\right) ,~\lambda \left(
t\right) $,~$\varepsilon \left( t\right) $ are determine in order the
equations to admit additional Lie symmetries.\ Finally, we apply the Lie
invariants to find similarity solutions for the generalized quantum
Zakharov-Kuznetsov equation.\newline
\newline
\newline

\end{abstract}

\section{Introduction}

Lie symmetry analysis is a powerful tool for the study of nonlinear
differential equations \cite{ibra,Bluman,Stephani,olver}. The pioneer
approach established by Sophus Lie is based on the determination of
one-parameter point transformations which leave invariant a given
differential equation. The existence of transformations which leave
invariant a differential equation indicates the existence of invariant
functions which can be used to write the corresponding differential into an
simpler form or into the form of another, well-known, differential equation.
The theory of symmetries provide a systematic way which has been applied the
last decades in \ a plethora of differential equations in all areas of
applied mathematics, we refer the reader to \cite%
{r1,r3,r5,mm01,mm02,m5,m6,m7,m8,m9,m10,m11} and references therein. For
other methods on the derivation of analytic solutions for differential
equations we refer the reader in \cite{sm1,sm2,sm3,sm4,bor} and references
therein.

In \cite{ovs}, Ovsiannikov classified all forms of the nonlinear heat
equation $u_{t}=\left( f\left( u\right) u_{x}\right) _{x}$ according to the
admitted Lie algebra. Since then, the classification problem has been widely
studied in the literature \cite{cl1,cl2,cl3,cl4,cl5,cl6,cl7,cl8,cl9}.\ In
this work we are interesting on the Lie symmetry analysis for the $2+1$
quantum Zakharov-Kuznetsov (qZK)%
\begin{equation}
u_{t}+uu_{z}+u_{zzz}+u_{xxz}=0.  \label{kz.02}
\end{equation}

The qZK equation describes weakly nonlinear ion--acoustic waves in the
presence of an uniform dense magnetic field. The quantum plasma has various
applications in \ many physical systems. Hence the qZK is an equation of
special interest. The Lie symmetry analysis for the Zakharov-Kuznetsov
equation, without the quantum terms, has been studied before in \cite{kzs1}.
The Lie symmetries for the fractional differential Zakharov-Kuznetsov\ (ZK)
were found in \cite{kzs2}, while for a modified ZK equation the symmetry
analysis was performed in \cite{kzs3}. As far as the $3+1$ qZK equation is
concerned, the Lie point symmetries were found for the first time in \cite%
{kzs4}. Finally the conservation laws for the qZK were constructed for the
first time in \cite{kzs5}.

In this work we extend our analysis and inspired by \cite{blaha} we consider
the generalized $2+1$ qZK equation
\begin{equation}
u_{t}+f\left( u\right) u_{z}+u_{zzz}+u_{xxz}=0,  \label{kz.03}
\end{equation}%
where $f\left( u\right) $ is an arbitrary function. Function $f\left(
u\right) $ is determined by the group properties of the differential
equation (\ref{kz.03}) as established by Ovsiannikov.

In addition we consider the $2+1~$qZK equation with time-varying
coefficients defined as \cite{ss1}
\begin{equation}
u_{t}+\delta \left( t\right) uu_{z}+\lambda \left( t\right)
u_{zzz}+\varepsilon \left( t\right) u_{xxz}=0.  \label{kz.04}
\end{equation}%
Again the time-varying coefficients are constrained according to the
admitted Lie symmetries. The plan of the paper is as follows.

In Section \ref{sec3} we present the basic properties and definitions for
the theory of symmetries for differential equations. The Lie point
symmetries for the $2+1$ qZK equation \ are determined in Section \ref{sec3}%
. We find that the $2+1$ qZK equation admits five Lie point symmetries. The
commutators and the adjoint representation of the admitted Lie symmetries
are calculated and are used to write the one-dimensional optimal system. The
symmetry vectors are used to define similarity transformations and to write
closed-form solutions. Specifically, the similarity transformations are used
to reduce the number of independent variables in the given differential
equation. By applying two similarity transformations we end with an ordinary
differential equation. We show that periodic solutions which belong to the
family to travelling-wave solutions exist. In\ Section \ref{sec4} we present
the complete classification scheme for the generalized $2+1$ qZK equation (%
\ref{kz.03}). The results are given in a proposition and a table. As far as
the time-dependent $2+1$ qZK equation (\ref{kz.04}) is concerned the Lie
point symmetries are studied in Section \ref{sec5}. Finally, in Section \ref%
{sec6} we summarize our results and we draw our conclusions.

\section{Preliminaries}

\label{sec2}

In this Section we present the basic properties and definitions for the
theory of Lie symmetries of differential equations. Consider the function $%
\Phi $ which describes the map of an one-parameter point transformation such
as $\Phi \left( u\left( t,x^{i}\right) \right) =u\left( t,x^{i}\right) \ $%
with infinitesimal transformation%
\begin{eqnarray}
t^{\prime } &=&t^{i}+\varepsilon \xi \left( t,x^{i},u\right)  \label{sv.12}
\\
x^{i\prime } &=&x^{i}+\varepsilon \xi ^{i}\left( t,x^{i},u\right) \\
u^{\prime } &=&u+\varepsilon \eta \left( t,x^{i},u\right)  \label{sv.13}
\end{eqnarray}%
and generator
\begin{equation}
X=\frac{\partial t^{\prime }}{\partial \varepsilon }\partial _{t}+\frac{%
\partial x^{\prime }}{\partial \varepsilon }\partial _{x}+\frac{\partial u}{%
\partial \varepsilon }\partial _{u},  \label{sv.16}
\end{equation}%
where~{$\varepsilon $ is the parameter of smallness}; $x^{i}=\left(
x,z\right) $, where $u\left( t,x^{i}\right) $ is the dependent function and $%
\left( t,x,z\right) $ are the independent variables.

Let $u\left( t,x^{i}\right) $ be a solution for the differential equation $%
\mathcal{H}\left( u,u_{,t},u_{,x}...\right) =0.$ Therefore under the
one-parameter map $\Phi $, function $u^{\prime }\left( x^{i\prime }\right)
=\Phi \left( u\left( x^{i}\right) \right) $ is a solution for the
differential equation $\mathcal{H}=0$, if and only if the differential
equation is also invariant under the action of the map, $\Phi $, that is,
the following condition holds
\begin{equation}
\Phi \left( \mathcal{H}\left( u,u_{t},u_{x}...\right) \right) =0.
\end{equation}

For every map $\Phi $ in which the latter condition holds it means that the
generator~$X$ is a Lie point symmetry for the differential equation while
\begin{equation}
\mathbf{X}^{\left[ n\right] }\left( \mathcal{H}\right) =0  \label{sv.17}
\end{equation}%
holds, where $\mathbf{X}^{\left[ n\right] }$ describes the $n^{th}$%
prolongation/extension of the symmetry vector in the jet-space of variables,
$\left\{ t,x^{i},u,u_{t},u_{i},u_{ij},...\right\} $.

The importance of the existence of a Lie symmetry for a given differential
equation is that from the associated Lagrange's system,%
\begin{equation}
\frac{dt}{\xi ^{t}}=\frac{dx^{i}}{\xi ^{i}}=\frac{du}{\eta },
\end{equation}%
invariants,~$U^{\left[ 0\right] }\left( t,x^{i},u\right) $ are able to be
determined which can be used to reduce the number of the independent
variables of the differential equation and lead to the construction of
similarity solutions. As far as partial differential equations are
concerned, the application of the Lie invariants reduces the number of the
independent variables. On the other hand, in the case of ordinary
differential equations the Lie invariants are applied to reduce the order
for the differential equation.

The admitted symmetry vectors of a given set of differential equations
constitute a closed-group known as a Lie group. \ The main application of
the Lie symmetries is the determination of solutions known as similarity
solutions and follow from the application of the Lie invariants in the
differential equations. However, in order to classify all the possible
similarity transformations and solutions the one-dimensional optimal system
should be calculated \cite{olver}.

Assume the $n$-dimensional Lie algebra $G_{n}$ with elements $\left\{
X_{1},~X_{2},~...~X_{n}\right\} ~$and structure constants $C_{jk}^{i}$. We
define the two symmetry vectors
\begin{equation}
Z=\sum_{i=1}^{n}a_{i}X_{i}~,~W=\sum_{i=1}^{n}b_{i}X_{i}~,~\text{\ }%
a_{i},~b_{i}\text{ are constants.}  \label{sw.04}
\end{equation}%
and we define the operator
\begin{equation}
Ad\left( \exp \left( \epsilon X_{i}\right) \right) X_{j}=X_{j}-\epsilon
\left[ X_{i},X_{j}\right] +\frac{1}{2}\epsilon ^{2}\left[ X_{i},\left[
X_{i},X_{j}\right] \right] +...  \label{sw.07}
\end{equation}%
known as the adjoint representation, in which $\left[ X_{i},X_{j}\right] $
is the Lie Bracket.

We say that the vectors $Z$ and $W$ are equivalent if and only if \cite%
{olver}
\begin{equation}
\mathbf{W}=\sum_{j=i}^{n}Ad\left( \exp \left( \epsilon _{i}X_{i}\right)
\right) \mathbf{Z}  \label{sw.05}
\end{equation}%
or
\begin{equation}
W=cZ~,~c=const\text{ that is }b_{i}=ca_{i}\text{.}  \label{sw.06}
\end{equation}

The one-dimensional subalgebras of $G_{n}$ which are not related through the
adjoint representation form the one-dimensional optimal system. The
determination of the one-dimensional system it is essential in order to
perform a complete classification of all the possible similarity
transformations and solutions.

\section{Symmetry analysis for the qZK}

\label{sec3}

For the qZK equation (\ref{kz.02}) the application of the Lie theory
provides that qZK~admits as Lie symmetries the elements of the five
dimensional Lie algebra
\begin{eqnarray*}
X_{1} &=&\partial _{t}~,~X_{2}=\partial _{x}~,~X_{3}=\partial _{z}~, \\
X_{4} &=&t\partial _{z}+\partial _{u}~,~X_{5}=3t\partial _{t}+x\partial
_{x}+z\partial _{z}.
\end{eqnarray*}%
The commutators and the adjoint representation for the admitted Lie
symmetries are presented in Tables \ref{tab1} and \ref{tab2} respectively.
We observe that the Lie symmeties form the $A_{4,2}\oplus A_{1}$ Lie algebra
in the Morozov-Mubarakzyanov classification \cite{mm1,mm2,mm3,mm4}

\begin{table}[tbp] \centering%
\caption{Commutator table for the admitted Lie point symmetries of the qKZ
equation}%
\begin{tabular}{cccccc}
\hline\hline
$\left[ X_{i},X_{j}\right] $ & $\mathbf{X}_{1}$ & $\mathbf{X}_{2}$ & $%
\mathbf{X}_{3}$ & $\mathbf{X}_{4}$ & $\mathbf{X}_{5}$ \\ \hline
$\mathbf{X}_{1}$ & $0$ & $0$ & $0$ & $X_{3}$ & $3X_{1}$ \\
$\mathbf{X}_{2}$ & $0$ & $0$ & $0$ & $0$ & $X_{2}$ \\
$\mathbf{X}_{3}$ & $0$ & $0$ & $0$ & $0$ & $X_{3}$ \\
$\mathbf{X}_{4}$ & $-X_{3}$ & $0$ & $0$ & $0$ & $-2X_{4}$ \\
$\mathbf{X}_{5}$ & $-3X_{1}$ & $-X_{2}$ & $-X_{3}$ & $2X_{4}$ & $0$ \\
\hline\hline
\end{tabular}%
\label{tab1}%
\end{table}%

\begin{table}[tbp] \centering%
\caption{Adjoint representation for the admitted Lie point symmetries of the
qKZ equation}%
\begin{tabular}{cccccc}
\hline\hline
$Ad\left( e^{\left( \varepsilon \mathbf{X}_{i}\right) }\right) \mathbf{X}%
_{j} $ & $\mathbf{X}_{1}$ & $\mathbf{X}_{2}$ & $\mathbf{X}_{3}$ & $\mathbf{X}%
_{4}$ & $\mathbf{X}_{5}$ \\ \hline
$\mathbf{X}_{1}$ & $X_{1}$ & $X_{2}$ & $X_{3}$ & $X_{4}-\varepsilon X_{3}$ &
$X_{5}-3\varepsilon X_{1}$ \\
$\mathbf{X}_{2}$ & $X_{1}$ & $X_{2}$ & $X_{3}$ & $X_{4}$ & $%
X_{5}-\varepsilon X_{2}$ \\
$\mathbf{X}_{3}$ & $X_{1}$ & $X_{2}$ & $X_{3}$ & $X_{4}$ & $%
X_{5}-\varepsilon X_{3}$ \\
$\mathbf{X}_{4}$ & $X_{1}+\varepsilon X_{3}$ & $X_{2}$ & $X_{3}$ & $X_{4}$ &
$X_{5}+2\varepsilon X_{4}$ \\
$\mathbf{X}_{5}$ & $e^{3\varepsilon }X_{1}$ & $e^{\varepsilon }X_{2}$ & $%
e^{\varepsilon }X_{3}$ & $e^{-2\varepsilon }X_{4}$ & $X_{5}$ \\ \hline\hline
\end{tabular}%
\label{tab2}%
\end{table}%

The one-dimensional optimal system consists of the following vector fields%
\begin{eqnarray*}
&&\left\{ X_{1}\right\} ~,~\left\{ X_{2}\right\} ~,~\left\{ X_{3}\right\}
~,~\left\{ X_{4}\right\} ~,~\left\{ X_{5}\right\} , \\
&&\left\{ X_{1}+\alpha X_{2}\right\} ~,~\left\{ X_{1}+\alpha X_{3}\right\}
~,~\left\{ X_{1}+\alpha X_{4}\right\} , \\
&&\left\{ X_{2}+\alpha X_{3}\right\} ~,~\left\{ X_{2}+\alpha X_{4}\right\}
~,~\left\{ X_{3}+\alpha X_{4}\right\} , \\
&&\left\{ X_{1}+\alpha X_{2}+\beta X_{4}\right\} ~,~\left\{ X_{2}+\alpha
X_{3}+\beta X_{4}\right\} ~, \\
&&\left\{ X_{1}+\alpha X_{2}+\beta X_{3}\right\} ~.
\end{eqnarray*}

We proceed with our analysis by applying the Lie symmetry vectors in order
to reduce the partial differential equation (\ref{kz.02}) into an ordinary
differential equation. Indeed, in order to perform such reduction we should
apply Lie point symmetries to perform the reduction process. Some
closed-form similarity solutions are presented.

\subsection{Similarity transformations}

We proceed by presenting the similarity transformations which follow by the
two-dimensional Lie algebras $\left\{ X_{4},X_{5}\right\} $~, $\left\{
X_{1}+\beta X_{2},X_{1}+\gamma X_{3}\right\} $. The solutions that we
present are those for which $u$ is function of all the variables $\left\{
t,x,z\right\} $.

\subsubsection{Solution $\left\{ X_{4},X_{5}\right\} $}

By using the Lie symmetry vectors $\left\{ X_{4},X_{5}\right\} $ we end to
the following ordinary differential equation $U_{\zeta }\zeta -U=0$, where $%
\zeta =xt^{-\frac{1}{3}}$ and $u\left( t,x,z\right) =\frac{z}{t}+U\left(
\zeta \right) t^{-\frac{2}{3}}$. Therefore the similarity solution is
derived to be
\begin{equation}
u\left( t,x,z\right) =\frac{z+x}{t}.
\end{equation}

\subsubsection{Solution $\left\{ X_{1}+\protect\beta X_{2},X_{1}+\protect%
\gamma X_{3}\right\} $}

Reduction with the symmetry vectors $\left\{ X_{1}+\beta X_{2},X_{1}+\gamma
X_{3}\right\} $ gives the travelling-wave solution $u=U\left( y\right)
~,~y=\beta z-\gamma t+\gamma x$, where $U\left( y\right) $ satisfies the
following differential equation%
\begin{equation}
\left( \beta ^{2}+\gamma ^{2}\right) U_{yy}-\gamma \beta ^{2}U+\frac{\beta
^{2}}{2}U^{2}-U_{1}=0  \label{eq1}
\end{equation}%
or, equivalently,%
\begin{equation}
\frac{\left( \beta ^{2}+\gamma ^{2}\right) }{2}U_{y}^{2}-\frac{\gamma \beta
^{2}}{2}U^{2}+\frac{\beta ^{2}}{6}U^{3}-U_{1}y-U_{0}=0.
\end{equation}%
The latter equation can be integrated by quadratures. \

Equation (\ref{eq1}) can be written as
\begin{equation}
U_{y}=V~,~V_{y}=\gamma \beta ^{2}U-\frac{\beta ^{2}}{2}U^{2}+U_{1}.
\label{eq2}
\end{equation}%
System (\ref{eq2}) admits two stationary points, they are
\begin{equation*}
U_{\pm }^{A}=\gamma \pm \sqrt{\gamma ^{2}-\frac{2U_{1}}{\beta ^{2}}\text{.}}
\end{equation*}%
These points are real when $\gamma ^{2}\geq \frac{2U_{1}}{\beta ^{2}}$. \
Easily we find that $U_{-}^{+}$ is always a source while $U_{+}^{A}$ is
always a centre point and describes periodic solutions.

\section{Group classification for the generalized qZK}

\label{sec4}

For the generalized $2+1$ qZK equation we find that for arbitrary function $%
f\left( u\right) $ the admitted Lie symmetries are the $\left\{
X_{1},~X_{2},~X_{3}\right\} $. However, for other functional forms of $%
f\left( u\right) $ equation (\ref{kz.03}) admits additional Lie symmetry
vectors. Hence for the Lie symmetry classification of the generalized $2+1$
qZK equation the following proposition follows.

\begin{proposition}
The generalized $2+1$ qZK equation (\ref{kz.03}) for an arbitrary function $%
f\left( u\right) $ admits three Lie point symmetries which form an Abelian
Lie algebra,~$3A_{1}$. Furthermore, for $f\left( u\right) =u_{0}$ the
generalized $2+1$ qZK equation admits an infinite number of Lie point
symmetries with finite algebra the $A_{4,2}\oplus A_{1}$. For $f\left(
u\right) =u$ it admits five Lie point symmetries which form the $%
A_{4,2}\oplus A_{1}$ Lie algebra. Finally Moreover, for the following
functional forms of $f\left( u\right) $, that is, $f_{B}\left( u\right)
=u^{\mu }+u_{0}$,~$f_{C}\left( u\right) =u+\kappa u^{2}+u_{0}$,~$f_{D}\left(
u\right) =e^{\mu u}+u_{0}$ and $f_{E}\left( u\right) =\ln u+u_{0}$ $\ $the
generalized $2+1$ qZK equation is invariant under four-dimensional Lie
algebras as they are presented in Table \ref{tabf}.
\end{proposition}

\begin{table}[tbp] \centering%
\caption{Lie symmetries classification scheme for the generalized qKZ
equation.}%
\begin{tabular}{cccc}
\hline\hline
$\mathbf{f}\left( u\right) $ & \textbf{Lie Algebra} & \textrm{dim}$\mathbf{G}%
_{n}$ & \textbf{Elements of}\textrm{\ }$\mathbf{G}_{n}$ \\ \hline
Arbitrary & $3A_{1}$ & $3$ & $X_{1}~,~X_{2}~,~X_{3}$ \\
$u$ & $A_{4,2}\oplus A_{1}$ & $5$ & $X_{1}~,~X_{2}~,~X_{3}~,~X_{4}~,~X_{5}$
\\
$u_{0}$ & $A_{3,1}\oplus 2A_{1}$ & $5~\&~\infty $ & $%
X_{1}~,~X_{2}~,~X_{3}~,~X_{4}^{A}~,~X_{5}^{A}~,~X_{b}$ \\
$u^{\mu }+u_{0}$ & $A_{4,2}$ & $4$ & $X_{1}~,~X_{2}~,~X_{3}~,~X_{4}^{B}$ \\
$u+\kappa u^{2}+u_{0}$ & $A_{4,2}$ & $4$ & $%
X_{1}~,~X_{2}~,~X_{3}~,~X_{4}^{C} $ \\
$e^{\mu u}+u_{0}$ & $A_{4,2}$ & $4$ & $X_{1}~,~X_{2}~,~X_{3}~,~X_{4}^{D}$ \\
$\ln u+u_{0}$ & $A_{4,2}$ & $4$ & $X_{1}~,~X_{2}~,~X_{3}~,~X_{4}$ \\
\hline\hline
\end{tabular}%
\label{tabf}%
\end{table}%

\subsection{Case A:$~f\left( u\right) =u_{0}$}

For a constant function $f\left( u\right) =u_{0}$, where without loss of
generality we assume $u_{0}=1$, equation (\ref{kz.03}) admits as Lie
symmetries the vector fields
\begin{equation*}
X_{1}~,~X_{2}~,~X_{3}~,~X_{4}^{A}=u\partial _{u},~X_{5}^{A}=3t\partial
_{t}+x\partial _{x}+\left( 2t+z\right) \partial _{z}\,\ \text{and~}%
X_{b}^{A}=b\left( t,x,z\right) \partial _{u}
\end{equation*}%
in which $b$ is a solution of the original equation. The symmetry vectors $%
X_{4}^{A},~X_{5}^{A}$ indicate the linearity for the partial differential
equation. The commutators and the adjoint representation for the admitted
Lie symmetries are presented in Tables \ref{tab3} and \ref{tab4}
respectively. The Lie symmtries form the $A_{3,1}\oplus 2A_{1}~$Lie algebra.

We observe that the Lie point symmetries for this case form a different Lie
algebra from that of equation (\ref{kz.02}). Hence, the resulting
one-dimensional optimal system is determined to consist of the symmetry
vectors%
\begin{eqnarray*}
&&\left\{ X_{1}\right\} ~,~\left\{ X_{2}\right\} ~,~\left\{ X_{3}\right\}
~,~\left\{ X_{4}\right\} ~,~\left\{ X_{5}\right\} , \\
&&\left\{ X_{1}+\alpha X_{2}\right\} ~,~\left\{ X_{1}+\alpha X_{3}\right\}
~,~\left\{ X_{1}+\alpha X_{4}\right\} , \\
&&\left\{ X_{2}+\alpha X_{3}\right\} ~,~\left\{ X_{2}+\alpha X_{4}\right\}
~,~\left\{ X_{3}+\alpha X_{4}\right\} , \\
&&\left\{ X_{1}+\alpha X_{2}+\beta X_{4}\right\} ~,~\left\{ X_{2}+\alpha
X_{3}+\beta X_{4}\right\} ~, \\
&&\left\{ X_{1}+\alpha X_{2}+\beta X_{3}\right\} ~,~\left\{ X_{1}+\alpha
X_{2}+\beta X_{3}+\gamma X_{4}\right\} .
\end{eqnarray*}

\begin{table}[tbp] \centering%
\caption{Commutator table for the admitted Lie point symmetries of the
generalized qKZ equation with $f(u)=f_{0}$}%
\begin{tabular}{cccccc}
\hline\hline
$\left[ X_{i},X_{j}\right] $ & $\mathbf{X}_{1}$ & $\mathbf{X}_{2}$ & $%
\mathbf{X}_{3}$ & $\mathbf{X}_{4}^{A}$ & $\mathbf{X}_{5}^{A}$ \\ \hline
$\mathbf{X}_{1}$ & $0$ & $0$ & $0$ & $0$ & $3X_{1}+2X_{3}$ \\
$\mathbf{X}_{2}$ & $0$ & $0$ & $0$ & $0$ & $X_{2}$ \\
$\mathbf{X}_{3}$ & $0$ & $0$ & $0$ & $0$ & $X_{3}$ \\
$\mathbf{X}_{4}^{A}$ & $0$ & $0$ & $0$ & $0$ & $0$ \\
$\mathbf{X}_{5}^{A}$ & $-3X_{1}-2X_{3}$ & $-X_{2}$ & $-X_{3}$ & $0$ & $0$ \\
\hline\hline
\end{tabular}%
\label{tab3}%
\end{table}%

\begin{table}[tbp] \centering%
\caption{Adjoint representation for the admitted Lie point symmetries of the
generalized qKZ equation with $f(u)=f_{0}$}%
\begin{tabular}{cccccc}
\hline\hline
$Ad\left( e^{\left( \varepsilon \mathbf{X}_{i}\right) }\right) \mathbf{X}%
_{j} $ & $\mathbf{X}_{1}$ & $\mathbf{X}_{2}$ & $\mathbf{X}_{3}$ & $\mathbf{X}%
_{4}^{A}$ & $\mathbf{X}_{5}^{A}$ \\ \hline
$\mathbf{X}_{1}$ & $X_{1}$ & $X_{2}$ & $X_{3}$ & $X_{4}^{A}$ & $%
X_{5}^{A}-3\varepsilon X_{1}-2\varepsilon X_{3}$ \\
$\mathbf{X}_{2}$ & $X_{1}$ & $X_{2}$ & $X_{3}$ & $X_{4}^{A}$ & $%
X_{5}^{A}-\varepsilon X_{2}$ \\
$\mathbf{X}_{3}$ & $X_{1}$ & $X_{2}$ & $X_{3}$ & $X_{4}^{A}$ & $%
X_{5}^{A}-\varepsilon X_{3}$ \\
$\mathbf{X}_{4}^{A}$ & $X_{1}$ & $X_{2}$ & $X_{3}$ & $X_{4}^{A}$ & $%
X_{5}^{A} $ \\
$\mathbf{X}_{5}^{A}$ & $e^{3\varepsilon }\left( X_{1}+X_{3}\right)
-e^{\varepsilon }X_{3}$ & $e^{\varepsilon }X_{2}$ & $e^{\varepsilon }X_{3}$
& $X_{4}^{A}$ & $X_{5}^{A}$ \\ \hline\hline
\end{tabular}%
\label{tab4}%
\end{table}%

\subsection{Case B: $f\left( u\right) =u^{\protect\mu }+u_{0}$}

For $f\left( u\right) =u^{\mu }+u_{0}$ the admitted Lie point symmetries are%
\begin{equation*}
X_{1}~,~X_{2}~,~X_{3}~,~X_{4}^{B}=\left( 3t\partial _{t}+x\partial
_{x}+\left( z+2u_{0}t\right) \partial _{z}-\frac{u}{\mu }\partial
_{u}\right) .
\end{equation*}%
The commutators and the adjoint representation for the admitted
four-dimensional Lie algebra are presented in Tables \ref{tab3} and \ref%
{tab4} respectively. By using the results of these Tables we can calculate
easily the one-dimensional optimal system composed of the one-dimensional
Lie algebras%
\begin{eqnarray*}
&&\left\{ X_{1}\right\} ~,~\left\{ X_{2}\right\} ~,~\left\{ X_{3}\right\}
~,~\left\{ X_{4}^{B}\right\} ~, \\
&&\left\{ X_{1}+\alpha X_{2}\right\} ~,~\left\{ X_{1}+\alpha X_{3}\right\}
~,~\left\{ X_{1}+\beta X_{4}^{B}\right\} ~, \\
&&\left\{ X_{2}+\alpha X_{3}\right\} ~,~\left\{ X_{1}+\alpha X_{2}+\beta
X_{3}\right\} ,
\end{eqnarray*}%
while the Lie symmetries form the $A_{4,2}$ Lie algebra.

\begin{table}[tbp] \centering%
\caption{Commutator table for the admitted Lie point symmetries of the
generalized qKZ equation with $f(u)=u^{\mu}+u_{0}$}%
\begin{tabular}{ccccc}
\hline\hline
$\left[ X_{i},X_{j}\right] $ & $\mathbf{X}_{1}$ & $\mathbf{X}_{2}$ & $%
\mathbf{X}_{3}$ & $\mathbf{X}_{4}^{B}$ \\ \hline
$\mathbf{X}_{1}$ & $0$ & $0$ & $0$ & $3X_{1}+2u_{0}X_{3}$ \\
$\mathbf{X}_{2}$ & $0$ & $0$ & $0$ & $X_{2}$ \\
$\mathbf{X}_{3}$ & $0$ & $0$ & $0$ & $X_{3}$ \\
$\mathbf{X}_{4}^{B}$ & $-3X_{1}-2u_{0}X_{3}$ & $-X_{2}$ & $-X_{3}$ & $0$ \\
\hline\hline
\end{tabular}%
\label{tabb1}%
\end{table}%

\begin{table}[tbp] \centering%
\caption{Adjoint representation for the admitted Lie point symmetries of the
generalized qKZ equation with $f(u)=u^{\mu}+u_{0}$}%
\begin{tabular}{ccccc}
\hline\hline
$Ad\left( e^{\left( \varepsilon \mathbf{X}_{i}\right) }\right) \mathbf{X}%
_{j} $ & $\mathbf{X}_{1}$ & $\mathbf{X}_{2}$ & $\mathbf{X}_{3}$ & $\mathbf{X}%
_{4}^{B}$ \\ \hline
$\mathbf{X}_{1}$ & $X_{1}$ & $X_{2}$ & $X_{3}$ & $X_{4}^{B}-3\varepsilon
X_{1}-2u_{0}\varepsilon X_{3}$ \\
$\mathbf{X}_{2}$ & $X_{1}$ & $X_{2}$ & $X_{3}$ & $X_{4}^{B}-\varepsilon
X_{2} $ \\
$\mathbf{X}_{3}$ & $X_{1}$ & $X_{2}$ & $X_{3}$ & $X_{4}^{B}-\varepsilon
X_{3} $ \\
$\mathbf{X}_{4}^{B}$ & $e^{3\varepsilon }\left( X_{1}+u_{0}X_{3}\right)
-u_{0}e^{\varepsilon }X_{3}$ & $e^{\varepsilon }X_{2}$ & $e^{\varepsilon
}X_{3}$ & $X_{4}^{B}$ \\ \hline\hline
\end{tabular}%
\label{tabb2}%
\end{table}%

\subsection{Case C: $f\left( u\right) =u+\protect\kappa u^{2}+u_{0}$}

For $f\left( u\right) =u+\kappa u^{2}$ the admitted Lie point symmetries are%
\begin{equation*}
X_{1}~,~X_{2}~,~X_{3}~,~X_{4}^{C}=6\kappa t\partial _{t}+2\kappa x\partial
_{x}+\left( 2\kappa z-t+4u_{0}\kappa t\right) \partial _{z}-\left( 1+2\kappa
u\right) \partial _{u}.
\end{equation*}%
The commutators and the adjoint representation for these four-dimensional
Lie algebra are presented in Tables \ref{tabc1} and \ref{tabc2}. From these
two tables we observe that the admitted Lie algebra is the same as that of
case B, i.e. the $A_{4,2}$, however, in a different representation. Thus,
the one-dimensional optimal system is comprised of the same one-dimensional
Lie algebras as that of case B.

\begin{table}[tbp] \centering%
\caption{Commutator table for the admitted Lie point symmetries of the
generalized qKZ equation with $f(u)=u+\kappa u^{2}+u_{0}$}%
\begin{tabular}{ccccc}
\hline\hline
$\left[ X_{i},X_{j}\right] $ & $\mathbf{X}_{1}$ & $\mathbf{X}_{2}$ & $%
\mathbf{X}_{3}$ & $\mathbf{X}_{4}^{C}$ \\ \hline
$\mathbf{X}_{1}$ & $0$ & $0$ & $0$ & $6\kappa X_{1}+\left( 4\kappa
u_{0}-1\right) X_{3}$ \\
$\mathbf{X}_{2}$ & $0$ & $0$ & $0$ & $2\kappa X_{2}$ \\
$\mathbf{X}_{3}$ & $0$ & $0$ & $0$ & $2\kappa X_{3}$ \\
$\mathbf{X}_{4}^{B}$ & $-6\kappa X_{1}-\left( 4\kappa u_{0}-1\right) X_{3}$
& $-2\kappa X_{2}$ & $-2\kappa X_{3}$ & $0$ \\ \hline\hline
\end{tabular}%
\label{tabc1}%
\end{table}%

\begin{table}[tbp] \centering%
\caption{Adjoint representation for the admitted Lie point symmetries of the
generalized qKZ equation with $f(u)=u+\kappa u^{2}+u_{0}$}%
\begin{tabular}{ccccc}
\hline\hline
$Ad\left( e^{\left( \varepsilon \mathbf{X}_{i}\right) }\right) \mathbf{X}%
_{j} $ & $\mathbf{X}_{1}$ & $\mathbf{X}_{2}$ & $\mathbf{X}_{3}$ & $\mathbf{X}%
_{4}^{C}$ \\ \hline
$\mathbf{X}_{1}$ & $X_{1}$ & $X_{2}$ & $X_{3}$ & $X_{4}^{B}-6\kappa
\varepsilon X_{1}-\left( 4\kappa u_{0}-1\right) \varepsilon X_{3}$ \\
$\mathbf{X}_{2}$ & $X_{1}$ & $X_{2}$ & $X_{3}$ & $X_{4}^{B}-2\kappa
\varepsilon X_{2}$ \\
$\mathbf{X}_{3}$ & $X_{1}$ & $X_{2}$ & $X_{3}$ & $X_{4}^{B}-2\kappa
\varepsilon X_{3}$ \\
$\mathbf{X}_{4}^{B}$ & $e^{6\kappa \varepsilon }\left( X_{1}+\left( u_{0}-%
\frac{1}{4\kappa }\right) X_{3}\right) -\left( u_{0}-\frac{1}{4\kappa }%
\right) e^{2\varepsilon }X_{3}$ & $e^{\varepsilon }X_{2}$ & $e^{\varepsilon
}X_{3}$ & $X_{4}^{B}$ \\ \hline\hline
\end{tabular}%
\label{tabc2}%
\end{table}%

\subsection{Case D: \thinspace $f\left( u\right) =e^{\protect\mu u}+u_{0}$}

For $f\left( u\right) =e^{\mu u}+u_{0}$ the admitted Lie point symmetries are%
\begin{equation*}
X_{1}~,~X_{2}~,~X_{3}~,~X_{4}^{D}=\left( 3t\partial _{t}+x\partial
_{x}+\left( z+2u_{0}t\right) \partial _{z}-\frac{2}{\mu }\partial
_{u}\right) .
\end{equation*}

The commutators and the adjoint representation are exactly the same as those
of case B presented in Tables \ref{tabb1} and \ref{tabb2}. Therefore, the
one-dimensional system is composed of the same one-dimensional Lie algebras.

\subsection{Case E: \thinspace $f\left( u\right) =\ln u+u_{0}$}

For $f\left( u\right) =\ln u+u_{0}$ the admitted Lie point symmetries are%
\begin{equation*}
X_{1}~,~X_{2}~,~X_{3}~,~X_{4}=t\partial _{z}+u\partial _{u}.
\end{equation*}%
We observe that that this is the fourth dimensional sub-algebra of the
original equation. Therefore, the commutators and the adjoint representation
are given in Tables \ref{tab1} and \ref{tab2} respectively. Moreover, the
one-dimensional optimal system is that for the qZK equation (\ref{kz.02})
except that here the vector field is $X_{5}$.

\section{Lie symmetries for the time-varying coefficient qZK}

\label{sec5}

In this Section we extend our analysis by studying the Lie symmetries for
the time-varying $2+1$ qZK equation (\ref{kz.04}). Without loss of
generality we can select $\delta \left( t\right) =1$. That it can be seen
easily by change the time variable $t\rightarrow T\left( \tau \right) $ and
define new coefficient functions. Thus in the following we assume $\delta
\left( t\right) =1$.

We apply the Lie symmetry condition and we summarize the results in the
following proposition.

\begin{proposition}
The time-varying $2+1$ qZK equation (\ref{kz.04}), for which without loss of
generality we have assumed $\delta \left( t\right) =1$, for arbitrary
functions $\lambda \left( t\right) $ and $\varepsilon \left( t\right) $.
Equation (\ref{kz.04}) admits a three-dimensional Lie algebra comprising the
symmetry vectors $~X_{2}=\partial _{x}~,~X_{3}=\partial _{z}$ and~$%
X_{4}=t\partial _{z}+\partial _{u}$. However, when $B\left( t\right) =t^{p}$%
,~$C\left( t\right) =t^{q}$, an additional symmetry vector exists, namely, $%
X_{T}^{1}=t\partial _{t}-\frac{\left( p-3q-2\right) }{6}x\partial _{x}+\frac{%
p+1}{3}z\partial _{z}+\frac{p-2}{3}u\partial _{u}$, while, when $B\left(
t\right) =e^{pt}$ and $C\left( t\right) =e^{qt}$, the additional symmetry
vector is $X_{T}^{2}=\partial _{t}-\frac{p-3q}{6}x\partial _{x}+\frac{p}{3}%
z\partial _{z}+\frac{p}{3}u\partial _{u}$.
\end{proposition}

The proof of this proposition is omitted. As far as the nonzero commutators
of the Lie symmetries are concerned for the time-dependent qZK equation we
find
\begin{equation*}
\left[ X_{2},X_{T}^{1}\right] =-\frac{\left( p-3q-2\right) }{6}X_{2}~,~\left[
X_{3},X_{T}^{1}\right] =\frac{p+1}{3}X_{3}~,~\left[ X_{4},X_{T}^{1}\right]
=X_{4},
\end{equation*}%
and%
\begin{equation*}
\left[ X_{2},X_{T}^{2}\right] =-\frac{p-3q}{6}X_{2}~,~\left[ X_{3},X_{T}^{2}%
\right] =\frac{p}{3}X_{3}~,~\left[ X_{4},X_{T}^{1}\right] =-X_{2}+\frac{p}{3}%
X_{4}.
\end{equation*}

\section{Conclusions}

\label{sec6}

In this piece of work, we studied the algebraic properties of the $2+1$ qZK
equation. In particular we solved the classification problem for the partial
differential equation (\ref{kz.03}) by determining all the functional forms
of $f\left( u\right) $ for which the equation admits Lie symmetries. For an
arbitrary function $f\left( u\right) $ the differential equation admits
three Lie point symmetries, while for linear function $f\left( u\right) $
admits five nontrivial Lie point symmetries. Moreover, for the following
cases, $f_{B}\left( u\right) =u^{\mu }+u_{0}~,~f_{C}\left( u\right)
=u+\kappa u^{2}+u_{0}$,~$f_{D}\left( u\right) =e^{\mu u}+u_{0}$ and $%
f_{E}\left( u\right) =\ln u+u_{0}$, the differential equation admits four
Lie point symmetries which form the Lie Algebra $A_{4,2}$. The results are
summarized in Proposition 1.

In addition, we consider the time-varying equation (\ref{kz.03}) with
nonconstant coefficients, and we classified the time-dependent coefficients
according to the admitted Lie point symmetries. Indeed, in the general case
the equation admits three Lie point symmetries, However, for the two special
cases described by Proposition 2 additional symmetries follow.

For the linear function $f\left( u\right) \,=u$ we applied the Lie
invariants in order to define similarity transformations and to reduce the
differential equation to an ordinary differential equation. We were able to
find a scaling solution and to prove the existence of travelling-wave
solutions. We do not proceed with the investigation of travelling-wave
solutions for the general case of arbitrary function $f\left( u\right) $ for
equation (\ref{kz.03}).

For an arbitrary function $f\left( u\right) $ the application of the Lie
point symmetries $\left\{ X_{1}+\beta X_{2},X_{1}+\gamma X_{3}\right\} $
reduces equation (\ref{kz.03}) to the partial differential equation%
\begin{equation}
\frac{\left( \beta ^{2}+\gamma ^{2}\right) }{\beta ^{2}}U_{yyy}-\left(
\gamma -f\left( U\right) \right) U_{y}=0,  \label{eq.03}
\end{equation}%
where $u=U\left( y\right) $ and $y=\beta z-\gamma t+\gamma x$. The
third-order differential equation can be integrated easily as%
\begin{equation}
\left( \beta ^{2}+\gamma ^{2}\right) U_{yy}-\beta ^{2}\left( \gamma
U-F\left( U\right) \right) U-U_{1}=0~,~~f\left( U\right) =\frac{dF\left(
U\right) }{dU}  \label{eq.04}
\end{equation}%
or, equivalently,
\begin{equation}
U_{y}=V~,~V_{y}=\gamma \beta ^{2}U-\beta ^{2}F\left( U\right) +U_{1}.
\label{eq.05}
\end{equation}%
Therefore, in order for the equation to admit periodic solution it should
follow that~the latter system admits at least a stationary point $U_{P}$ in
which $\gamma \beta ^{2}U\,_{P}-\beta ^{2}F\left( U_{P}\right) +U_{1}$ and $%
f\left( U_{P}\right) >\gamma $.

This work contributes to the subject of the group properties of differential
equations and specifically of plasma physics differential equations. In a
future work we plan to investigate the derivation of conservation laws for
the generalized $2+1$ qZK equation.

\end{document}